\documentclass[12pt]{article}

\usepackage{fullpage}
\usepackage{CJKutf8} 
\usepackage[utf8]{inputenc}
\usepackage{setspace}
\usepackage{parskip}
\usepackage{titlesec}
\usepackage{placeins}
\usepackage{xcolor}
\usepackage{breakcites}
\usepackage{lineno}
\usepackage{hyphenat}
\usepackage[dvipsnames]{xcolor}
\usepackage[nomarkers,nolists]{endfloat}

\usepackage{geometry}

\date{}

\PassOptionsToPackage{hyphens}{url}
\usepackage[colorlinks = true,
            linkcolor = blue,
            urlcolor  = blue,
            citecolor = blue,
            anchorcolor = blue]{hyperref}            

\usepackage[numbers,super,sort&compress]{natbib}

\makeatletter
\renewcommand{\@biblabel}[1]{#1.}
\makeatother

\renewenvironment{abstract}
  {{\bfseries\noindent{\abstractname}\par\nobreak}\footnotesize}
  {\bigskip}

\titlespacing{\section}{0pt}{*3}{*1}
\titlespacing{\subsection}{0pt}{*2}{*0.5}
\titlespacing{\subsubsection}{0pt}{*1.5}{*0.5}

\usepackage{authblk}

\usepackage{listings}
\usepackage{xcolor}

\usepackage{graphicx}
\usepackage[font=normalsize,labelfont=bf,labelsep=period]{caption}
\usepackage[space]{grffile}
\usepackage{latexsym}
\usepackage{textcomp}
\usepackage{longtable}
\usepackage{tabulary}
\usepackage{booktabs,array,multirow}
\usepackage{amsfonts,amsmath,amssymb}
\providecommand\citet{\cite}
\providecommand\citep{\cite}

\newif\iflatexml\latexmlfalse

\AtBeginDocument{\DeclareGraphicsExtensions{.pdf,.PDF,.eps,.EPS,.png,.PNG,.tif,.TIF,.jpg,.JPG,.jpeg,.JPEG}}

\usepackage[english]{babel}
\usepackage{threeparttable}

\usepackage{siunitx}
\usepackage{pdflscape}

\begin{document}
\begin{CJK}{UTF8}{gbsn}



\title{Authentic Multinational Federated Time-to-Event Analyses Among People with HIV in Latin America
}

\author[1]{Kaixing Liu}
\author[1]{Zhuohui J. Liang}
\author[2]{Fabio Paredes}
\author[3]{Ronaldo I. Moreira}
\author[4]{Yanink Caro-Vega}
\author[5]{Jiayi Tong}
\author[6]{Zhuohang Li}
\author[7]{Carina Cesar}
\author[8]{Yong Chen}
\author[9]{Jessica L. Castilho}
\author[10]{Stephany N. Duda}
\author[1,6,10,11]{Bradley A. Malin}
\author[10]{Chao Yan\thanks{Co-corresponding author: \texttt{chao.yan.1@vumc.org}}}
\author[1,10]{Bryan E. Shepherd\thanks{Corresponding author: \texttt{bryan.shepherd@vanderbilt.edu}}}
\author[12]{On behalf of the Caribbean, Central, and
South America network for HIV epidemiology (CCASAnet)}
\affil[1]{Department of Biostatistics, Vanderbilt University Medical Center, Nashville, TN, USA}%
\affil[2]{Pontificia Universidad Católica de Chile, Santiago, Chile}
\affil[3]{Instituto Nacional de Infectologia Evandro Chagas, Fundação Oswaldo Cruz (INI-Fiocruz), Rio de Janeiro, Brazil}
\affil[4]{Departamento de Infectología, Clínica de Inmunoinfectología VIH, Instituto Nacional de Ciencias Médicas y Nutrición Salvador Zubirán, Tlalpan, Ciudad de México, México}
\affil[5]{Department of Biostatistics, Johns Hopkins University, Baltimore, MD, USA}
\affil[6]{Department of Computer Science, Vanderbilt University, Nashville, TN, USA}
\affil[7]{Research Department, Fundación Huésped, Buenos Aires, Argentina}
\affil[8]{Department of Biostatistics, Epidemiology, and Informatics, University of Pennsylvania, Philadelphia, PA, USA}
\affil[9]{Division of Infectious Diseases, Vanderbilt University Medical Center, Nashville, TN, USA}
\affil[10]{Department of Biomedical Informatics, Vanderbilt University Medical Center, Nashville, TN, USA}
\affil[11]{Department of Electrical and Computer Engineering, Vanderbilt University, Nashville, TN, USA}
\affil[12]{CCASAnet, USA}

\begingroup
\let\center\flushleft
\let\endcenter\endflushleft
\newgeometry{margin=1in,top=0.75in,bottom=0.75in}
\maketitle
\endgroup

\vspace{-2em}

{\footnotesize

\subsubsection*{Mailing Address of Corresponding Author}


2525 West End Avenue, Suite 1100\\
Nashville, TN 37203

\subsubsection*{Acknowledgements}
AI-based tools (e.g., Claude) were used to refine figure-generation code and to assist with structuring, documenting, and formatting the code released in the accompanying GitHub repository. All code was reviewed and verified by the authors.

\subsubsection*{Funding}
This work was supported in part by the NIH‐funded Caribbean, Central and South America network for HIV epidemiology (CCASAnet), a member cohort of the International epidemiologic Databases to Evaluate AIDS (IeDEA) (U01AI069923), K99LM01442801, R01MH139379, and the Tennessee Center for AIDS Research (P30 AI110527).

\subsubsection*{Disclaimer}

The content is solely the responsibility of the authors and does
not necessarily represent the official views of the NIH, the CCASAnet, or the Tennessee Center for AIDS Research.

\subsubsection*{Data Availability Statement}
The analysis code and the protocols for authentic federated implementation are available in the GitHub repository at \url{https://github.com/CharlieLiu220/federated-survival-analysis-CCASAnet}.

}
\restoregeometry
\clearpage


\newpage

\begin{abstract}

Multinational HIV cohort studies face regulatory barriers to cross-border sharing of individual participant data, limiting centralized pooled analyses.
Federated statistical methods, which exchange only aggregated information, offer a privacy-preserving alternative but have rarely been examined in real-world distributed environments for HIV research. Here, we evaluate the feasibility and analytical performance of a communication-efficient federated framework 
within the Caribbean, Central, and South America Network for HIV epidemiology.
Virologic failure and major regimen change after antiretroviral therapy initiation were analyzed as separate outcomes; for each, we estimated cumulative incidence functions (CIFs) and fit stratified cause-specific Cox proportional hazards models via a surrogate likelihood-based federated implementation. 
Each site imputed missing data, conducted local analysis, and shared only summary statistics according to a coordinated computation protocol. The federated approach exactly reproduced centralized CIFs and closely approximated centralized Cox regression estimates, outperforming conventional meta-analysis for both outcomes.
These findings demonstrate that authentic federated analysis is feasible for multinational HIV research and can yield results closely aligned with centralized analysis while preserving data privacy. Post-hoc feedback from local analysts, however, indicated that broader adoption will require managing the logistical and coordination overhead and ensuring harmonized data collection and quality control across sites. 

\end{abstract}


\section*{Introduction}
International collaboration is foundational to HIV epidemiology, where individual participant data (IPD) from diverse geographical sources are often pooled to enable statistical analyses that produce accurate and generalizable conclusions.\cite{egger2002} However, due to privacy concerns and an increasingly complex regulatory landscape,
international HIV data sharing has become more difficult, limiting the generation of scientific evidence and slowing downstream translational impact. \cite{xia2024paradigm,lalova2024euus,bentzen2021remove} One natural workaround is meta-analysis, which has been widely used to aggregate study-level summaries (e.g., coefficient estimates and standard errors) from site- or institution-specific analyses, without sharing IPD. Yet, meta-analysis can yield inaccurate results with unstable estimates and inflated variance, especially in settings with small sample sizes, rare outcomes, and heterogeneous populations. \cite{lu2025meta}

More recently, federated statistical analysis has emerged as a promising alternative that coordinates local analysis across multiple sites, fully or mostly recovering the centralized estimates while keeping IPD local. Compared to meta-analysis, federated 
analysis offers greater statistical accuracy by leveraging individual-level variation across sites, thereby reducing information loss and enabling more complex modeling of covariates and heterogeneity.\cite{lu2025meta} 

Federated statistical analysis generally encompasses two distinct approaches: 1) iterative methods, and 2) communication-efficient methods. For base statistical models that estimate with Fisher-Scoring or Newton-Raphson algorithms (e.g., logistic regression and Cox proportional hazards regression), an iterative federated method decomposes calculation of components required per iteration into site-specific tasks and then combines tasks by involving sites sharing intermediate results at each iteration until convergence. \cite{wu2012glore,lu2015webdisco} Such methods are immune to information loss in the sense that they can exactly reproduce the centralized analysis but at the cost of operational efficiency, namely necessitating many rounds of communication and coordination across sites. In contrast, communication-efficient methods involve only a few iterations of exchange of aggregated information but usually come with some degree of information loss. Built upon the surrogate likelihood method,\cite{jordan2019communication} ODAL\cite{duan2020odal} and ODAC\cite{duan2020odac} are two examples of communication-efficient federated algorithms for logistic regression and Cox regression, respectively. It is worth noting that for statistical models with closed-form solutions, e.g., least squares regression, a communication-efficient algorithm can also be lossless. \cite{chen2006regressioncubes} Taken together, communication-efficient federated algorithms \cite{duan2020odal,duan2020odac,luo2022odach,li2023distributed} that maintain sufficient accuracy present a more viable path forward than iterative methods,\cite{lu2015webdisco,wu2012glore} whose limited scalability and practical deployment challenges constrain their broader implementation.

Despite their statistical merits, federated algorithms have largely been tested in simulated multi-site settings, with a single analyst emulating the federated process using partitioned real-world data in the same computational environment.\cite{duan2020odal,duan2020odac,luo2022odach,li2023distributed} Such simulations, while valuable for evaluating statistical accuracy, inherently bypass the logistical, communication, and technical challenges of true cross-site collaboration. Moreover, most operate under an assumption of complete data, neglecting data missingness, an issue that is pervasive in practice. To date, authentic deployments of these algorithms in the HIV domain, where analysts at each participating site run the federated workflow on their own data in their local computational environments, remain exceedingly rare.

In this paper, we investigate the feasibility and the accuracy 
of implementing an operationally efficient federated stratified Cox regression for time-to-event outcomes in an authentic distributed environment using real-world data from a multinational HIV cohort. Specifically, we implement a communication-efficient federated algorithm, performing sub-analyses at sites in three countries separately and pooling information without sharing IPD, 
while addressing complexities arising in real data such as missing data. We demonstrate the similarity of results derived from the federated algorithms to those from centralized analyses. We also describe challenges with real-world implementation and lessons learned.

\section*{Methods}
\subsection*{Data}
We compared centralized analysis with federated analysis as well as meta-analysis using data from the Caribbean, Central and South America network for HIV epidemiology (CCASAnet). This network has established a shared repository of data from people with HIV (PWH) at sites in Latin America to study the regional HIV epidemic. \citep{mcgowan2007} An earlier CCASAnet study \cite{cesar2015incidence} investigated the incidence and risk factors of virologic failure and major regimen change following initiation of antiretroviral therapy (ART) between 2000 and 2014. With the onset of the ``Treat-All'' era in 2015 \cite{who2015} and the widespread use of modern first-line regimens such as those containing dolutegravir and bictegravir, there is interest in updating estimates and investigating risk factors for these two critical HIV endpoints.
\par
The CCASAnet sites contributing data to this specific study are in Brazil (Instituto Nacional de Infectologia, Evandro Chagas, Fundacao Oswaldo Cruz in Rio de Janeiro), Chile (Fundacion Arriaran in Santiago), and Mexico (Instituto Nacional de Ciencias Medicas y Nutricion Salvador Zubiran in Mexico City). For the remainder of this manuscript, we use country names to denote HIV data collection and analysis sites. Treatment-naive adults who were at least 18 years old at ART initiation and began their first ART regimen on or after January 1\textsuperscript{st}, 2015 were included in the analyses, while those with ART treatment prior to 2015 were excluded. For the virologic failure analyses, we required PWH to have at least one viral load measurement after ART initiation to be included. Following the earlier CCASAnet study, \cite{cesar2015incidence} the endpoints were defined separately as follows:
\par
An individual experienced virologic failure if one of the following was satisfied:
    \begin{enumerate}
        \item measurement of HIV-1 RNA (viral load; VL) never dropped below 200 copies/mL after 6 months of ART;
        \item VL measurement dropped below 200 copies/mL but then there were two consecutive values $>$200 copies/mL;
        \item VL measurement dropped below 200 copies/mL but then there was a single measurement $>$1000 copies/mL.
    \end{enumerate}
\par
An individual underwent a major regimen change if the backbone drug of the initial ART regimen (i.e., non-nucleoside reverse transcriptase inhibitor [NNRTI], (boosted) protease inhibitor [(b)PI], or integrase strand transfer inhibitor [INSTI]) was switched.
\par
We considered the time from ART initiation to 1) virologic failure and  2) major regimen change separately. Our goals were to estimate the time-varying probabilities of each event and to  identify associated risk factors. Covariates considered for both outcomes include sex, age at ART initiation, probable route of HIV acquisition, type of initial ART regimen defined by the backbone ART class, baseline measurements of VL and CD4 count, calendar year of ART initiation, and baseline clinical AIDS status. The supplemental material contains details defining these covariates. This study was approved by the VUMC Institutional Review Board, number 060284.
\par

\subsection*{Analyses}
We performed four types of analyses: 1) centralized, 2) meta-analyses, 3) simulated, and 4) authentic federated analyses. We provide a brief overview of these different types of analyses, with further details in the supplemental material.

\subsubsection*{Centralized Analyses}

IPD from each CCASAnet site were sent to the data coordinating center at Vanderbilt University Medical Center (VUMC). Hence, centralized analyses were able to be performed at VUMC, the results of which served as the gold standard for subsequent evaluations.

Overall and site-specific cumulative incidence functions (CIF) for both outcomes were estimated with the Aalen-Johansen estimator, \cite{aalen1978empirical} treating death as a competing event. Cause-specific stratified Cox proportional hazards models \cite{cox1972} were employed to evaluate the association between risk factors and hazards of each event in adjusted analyses. Cox models were stratified by site, allowing the underlying hazard functions to vary between sites but assuming common hazard ratios for covariates across sites. We carried out both complete-case analyses, in which only observations with complete covariate and outcome data were included, and analyses that addressed missing data with multiple imputation using chained equations (MICE). Specifically, missing data were imputed 10 times locally (i.e., independently for each site). For the centralized adjusted analyses, each imputed full dataset was formed by combining each locally imputed dataset across sites. Analyses were performed on the combined full datasets while Rubin's rules were used to aggregate results across imputation replications.\cite{rubin1987}

\subsubsection*{Meta-Analyses}

For the meta-analysis, each site fit an unstratified Cox proportional hazards model locally. To synthesize the results from each site, a fixed-effects model was imposed on each coefficient, where the fitted coefficients were assumed to follow a normal distribution.\cite{sutton2000} Consequently, point estimates for each coefficient were obtained as the weighted averages of site-specific coefficient estimates; standard errors were also estimated. Missing data were imputed 10 times locally in a manner identical to that for the centralized analyses. Local analyses based on multiply imputed datasets were combined using Rubin's rules for each site, prior to meta-analysis.

\subsubsection*{Federated Analyses}

The federated adjusted analyses were performed using the communication-efficient and privacy-preserving algorithm, ODACH.\cite{luo2022odach} ODACH is based on the surrogate likelihood framework\cite{jordan2019communication} and follows a divide-and-aggregate principle, which is possible because the global partial likelihood of the stratified Cox model can be decomposed into site-specific components. Another federated algorithm for the stratified Cox model was also examined, which modifies the Newton-Raphson algorithm used for estimating regression coefficients via extra Taylor series expansions. \cite{li2023distributed} Since this algorithm was only successfully executed for the complete-case analysis for virologic failure, where it performed uniformly worse than ODACH (Table S2 in the supplemental material), and it was not ready for implementation in an authentic distributed environment, we focused on ODACH for simulated and authentic federated analyses. Now we present the procedures of ODACH.
\par
Let $\mathcal D_k$ represent the IPD locally available at site $k$, for $k=1,\ldots, K$, and define $\mathcal D=\{\mathcal D_k\}_{k=1}^K$, as the pooled IPD across sites.  In short, ODACH works as follows:\cite{luo2022odach}
\begin{enumerate}
    \item Initially, each site performs a local analysis via unstratified Cox proportional hazards regression and shares estimated regression coefficients and their variance-covariance matrix, $(\hat{\boldsymbol \beta}_k, \hat{\boldsymbol \Sigma}_k)$, with the lead site (e.g., $k=1$) (initialization);
    \item Next, the lead site calculates (according to Equation~\ref{eq:initial}) and broadcasts $\bar {\boldsymbol \beta}$ (initialization); 
    \begin{equation}\label{eq:initial}
\bar {\boldsymbol \beta}=(\sum_{k=1}^K \hat{\boldsymbol \Sigma}_k^{-1})^{-1}\sum_{k=1}^K \hat{\boldsymbol \Sigma}_k^{-1}\hat{\boldsymbol \beta}_k        
    \end{equation}
    \item Each site then calculates the score function and the Hessian matrix, evaluated at $\bar {\boldsymbol \beta}$, denoted as $\nabla L_k(\bar {\boldsymbol \beta};\mathcal D_k)$ and $\nabla^2 L_k(\bar {\boldsymbol \beta};\mathcal D_k)$, respectively, and shares these with the lead site (derivation);
    \item Afterwards, the lead site computes $\nabla L(\bar {\boldsymbol \beta};\mathcal D)$ by Equation~\ref{eq:score} and $\nabla ^2L(\bar {\boldsymbol \beta};\mathcal D)$ by Equation~\ref{eq:hessian},
\begin{equation}\label{eq:score}
\nabla L(\bar {\boldsymbol \beta};\mathcal D)=\sum_{k=1}^K \nabla L_k(\bar {\boldsymbol \beta};\mathcal D_k)    
\end{equation}
\begin{equation}\label{eq:hessian}
\nabla ^2L(\bar {\boldsymbol \beta};\mathcal D)=\sum_{k=1}^K \nabla^2 L_k(\bar {\boldsymbol \beta};\mathcal D_k)    
\end{equation}
    and then updates its log partial likelihood through Equation~\ref{eq:surrogate}.
\begin{equation}\label{eq:surrogate}
\begin{split}
  \tilde{L}(\boldsymbol {\beta}; \mathcal{D}_1)
    ={}& L_1(\boldsymbol {\beta}; \mathcal{D}_1)
        + \bigl(\nabla L(\bar{\boldsymbol {\beta}}; \mathcal{D})
                - \nabla L_1(\bar{\boldsymbol {\beta}}; \mathcal{D}_1)\bigr)^\top \boldsymbol{\beta} \\
       & + \tfrac{1}{2}(\boldsymbol {\beta} - \bar{\boldsymbol {\beta}})^\top
           \bigl(\nabla^2 L(\bar{\boldsymbol{\beta}}; \mathcal{D})
                 - \nabla^2 L_1(\bar{\boldsymbol {\beta}}; \mathcal{D}_1)\bigr)
           (\boldsymbol {\beta} - \bar{\boldsymbol{\beta}})
\end{split}
\end{equation} 
    
    Finally, estimation is completed at the lead site by maximizing $\tilde L(\boldsymbol \beta;\mathcal D_1)$ with respect to $\boldsymbol \beta$ (estimation).
\end{enumerate}

Any general and appropriate optimization program beyond a Newton-Raphson algorithm can be leveraged to conduct the estimation based on $\tilde L(\boldsymbol \beta;\mathcal D_1)$ at the lead site. 
In theory, each site can construct its own $\tilde L(\boldsymbol \beta;\mathcal D_k)$ by letting each site broadcast rather than share in steps 1 and 3 above, and estimates based on $\tilde L(\boldsymbol \beta;\mathcal D_k)$ across sites can then be combined via meta-analysis to produce final estimates. In practice, however, the participating site with the largest sample size is recommended to be the lead site.\cite{luo2022odach} Here, $K=3$ and Chile, with the largest sample size (Table \ref{table1}), was selected as the lead site.

Since CIFs can be estimated without numerical optimization, a federated approach to estimating CIFs can be information lossless. In particular, the federated approach obtains exactly the same estimates as those in the centralized analysis because those estimates can be perfectly reconstructed by each site sharing summarized cumulative counts to the lead site without sharing IPD. Details are in the supplemental material.

\emph{Simulated federated analyses} were performed by an analyst at VUMC with access to the entire data, who simulated a federated analysis by performing site-specific tasks of federated analyses separately and then combined information across sites, all within a single computational environment. Simulated federated analyses of this type are commonly employed to illustrate the performance of newly proposed federated methods.

\begin{figure}[!ht]
\centering
\includegraphics[width=1.00\columnwidth]{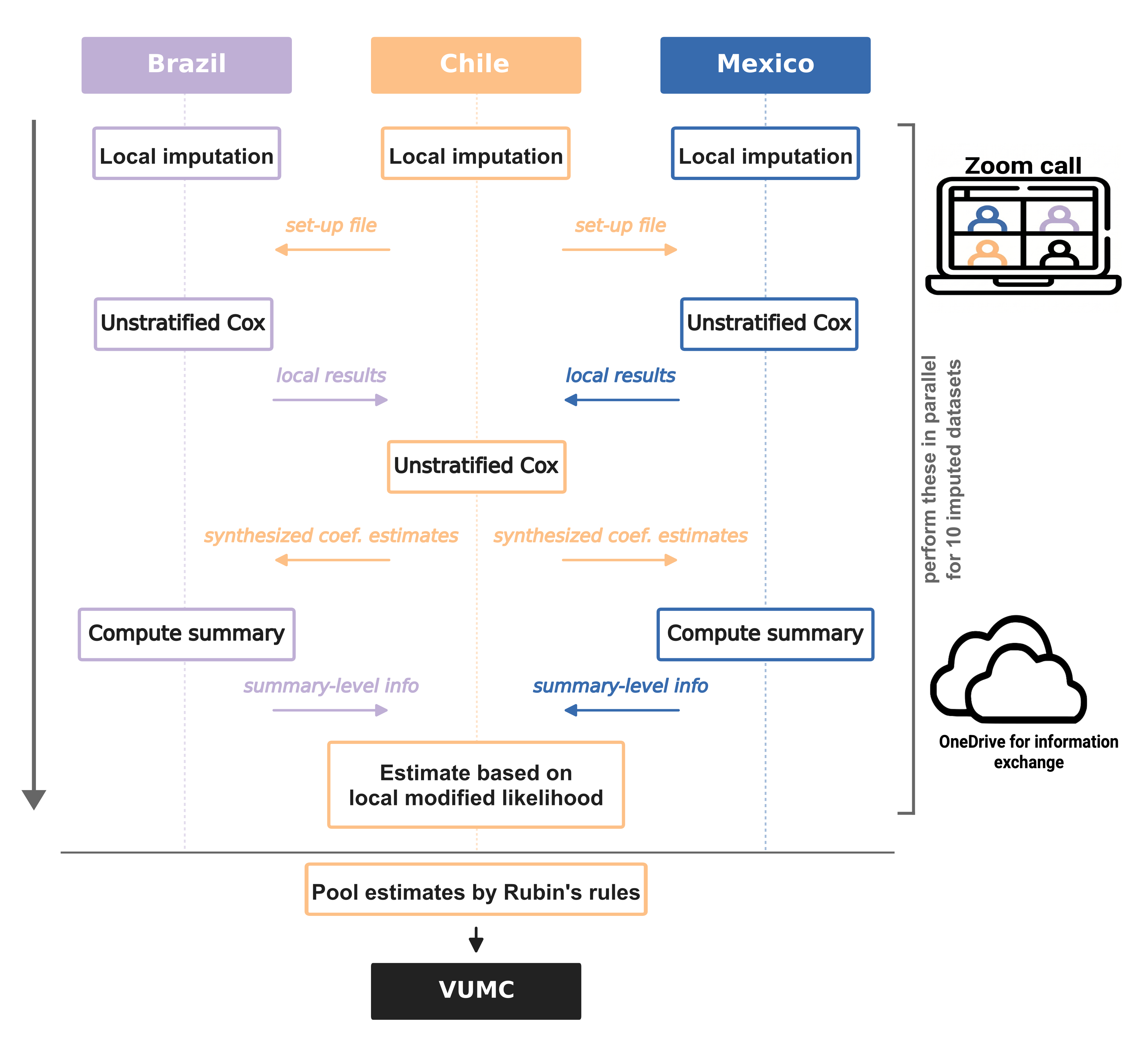}
\vspace{-1em}
\caption{Process of authentic federated analyses by ODACH with multiple imputation: Chile was selected as the lead site while Brazil and Mexico were participating site; coef. is abbreviated for coefficient; info refers to information; Zoom application was used for holding video conferences; VUMC-authorized OneDrive was used to download/upload files for information exchange.
}
\label{figure0}
\end{figure}

\emph{Authentic federated analyses} were performed by analysts at local sites, with no sharing of IPD across sites. Each of the three local analysts is a statistician and/or data manager at their site with graduate-level training. Source code and site-specific protocols for executing federated analyses were developed at VUMC and tested as part of the simulated federated analyses. The code, along with the protocols, were then sent to the local sites. Figure~\ref{figure0} shows the process of authentic federated analyses by ODACH with multiple imputation. Prior to conducting authentic federated analyses, video calls were held with each site separately to ensure that the required software was correctly installed, that the site-specific code executed without errors, and that the local analysts were familiar with their protocols. Since ODACH involves two rounds of information exchange between a lead site (Chile) and participating sites (Brazil and Mexico) and local execution of the code should strictly follow the protocols, another two video calls were scheduled to expedite the implementation and to make sure the implementation would be successful. A structured post-hoc review via a REDCap survey was conducted with each site's analyst to gather information on their experience participating in the federated analysis.

\subsubsection*{Evaluation} 
Comparisons between the federated, meta-, and centralized analyses were displayed using point estimates and 95\% confidence intervals (CI) and were made analytically via Wasserstein distance,\cite{panaretos2019statistical} both on the log-hazard ratio scale. Specifically, the distribution of the point estimate for each regression coefficient was assumed to follow a normal distribution with mean and standard deviation given by $(\hat \beta_j, \hat\sigma_j)$ for the federated learning and the meta-analysis and $(\hat \beta_j^*, \hat\sigma_j^*)$ for the centralized analysis; the Wasserstein distance to the centralized analysis was given by Equation~\ref{eq:wd}. As a dissimilarity metric, a smaller value of Wasserstein distance indicates more similarity.

\begin{equation}\label{eq:wd}
    d_j=\sqrt {(\hat \beta_j-\hat \beta_j^*)^2+(\hat \sigma_j-\hat \sigma_j^*)^2}
\end{equation}

\subsubsection*{Software and Implementation}
All analyses were performed using R statistical software, version 4.4.1. The software for executing ODACH is available as the package \{pda\} version 1.2.5\cite{rpackagepda} in R. In authentic federated analyses, analysts at the local sites executed site-specific R code and shared intermediate aggregated information via VUMC-authorized OneDrive during video-conference calls, as shown in Figure~\ref{figure0}. Refer to the supplemental material for details.

\par
\par



\section*{Results}
\subsection*{Characteristics of the Study Cohort}
A total of 5,541 PWH were included in this study (Brazil, 1,928; Chile, 3,220; Mexico, 393). Table \ref{table1} shows baseline information at ART initiation and the frequency of events across sites. The median age was 31-32 years and the majority were men who had sex with men (MSM). Most individuals (77-83\%) did not have clinical AIDS at ART initiation and median CD4 count ranged from 245 to 339 cells/mm$^3$. Missingness was primarily due to CD4 count and VL measurements at baseline. Individuals tended to have started ART more recently in Brazil and Chile, while NNRTI-based initial ART regimens were more often prescribed in Mexico.
All individuals were included in the analyses for major regimen change, while for virologic failure, the pooled sample size was reduced to a subset of 4,670 (Brazil, 1,802; Chile, 2,488; Mexico, 380) due to missing VL measurements after ART initiation. The proportion of events at each site for either outcome ranged from 7.7\% to 15.5\% except that in Mexico, more than half of the analysis sample experienced a major regimen change.

\subsection*{Cumulative Incidence of Events}
In the combined cohort, the estimated cumulative incidence for virologic failure five years after ART initiation was 0.177 (95\% CI, 0.160–0.196), and 0.316 (95\% CI, 0.296–0.338) for major regimen change. There was considerable heterogeneity in the estimated cumulative incidences between sites.

\begin{figure}[!ht]
\centering
\includegraphics[width=1.05\columnwidth]{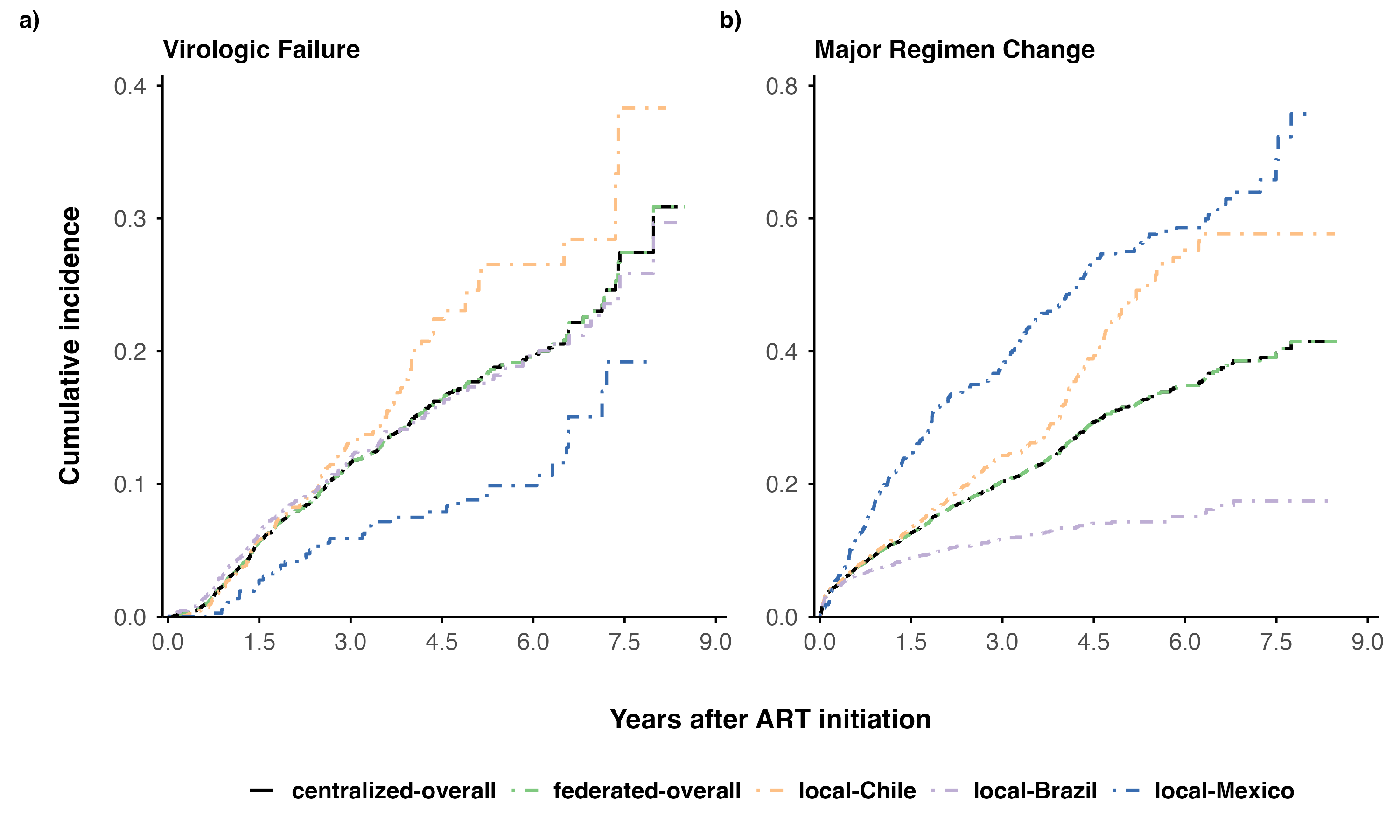}
\vspace{-2.5em}
\caption{Overall and site-specific cumulative incidence functions of virologic failure and major regimen change: centralized/federated-overall refers to the overall CIF obtained by the centralized analysis/the authentic federated approach; local-Chile/Brazil/Mexico represents the site-specific CIF based on IPD at each site.
}
\label{figure1}
\end{figure}

The plots of CIFs (Figure~\ref{figure1}) illustrate the overall and local trends for each outcome, accounting for death as a competing event. Importantly, the two overall CIFs obtained from the centralized analysis and the authentic federated analysis with respect to the same outcome completely overlapped, demonstrating the lossless property of the federated approach to calculating CIFs. While the confidence intervals are not shown, they are also perfectly overlapped between the centralized and the federated analyses.


\subsection*{Adjusted Analyses}
The centralized Cox regression analyses produced consistent results for both complete-case (Table S1) and multiply imputed analyses (Table~\ref{table2}). Individuals who were male, older, or had higher baseline CD4 cell count had lower hazards of virologic failure, while individuals who were diagnosed with clinical AIDS at baseline, initiated ART in later calendar years, or started a PI-based ART regimen (as compared to a NNRTI-based regimen) had higher hazards of virologic failure. Being male or free of AIDS was also associated with a lower hazard of having a major regimen change. Starting a PI-based regimen was associated with a greater hazard of experiencing a major regimen change, while individuals starting an INSTI-based regimen were less likely to experience a major regimen change. 

\begin{figure}[!ht]
\centering
\includegraphics[width=1.05\columnwidth]{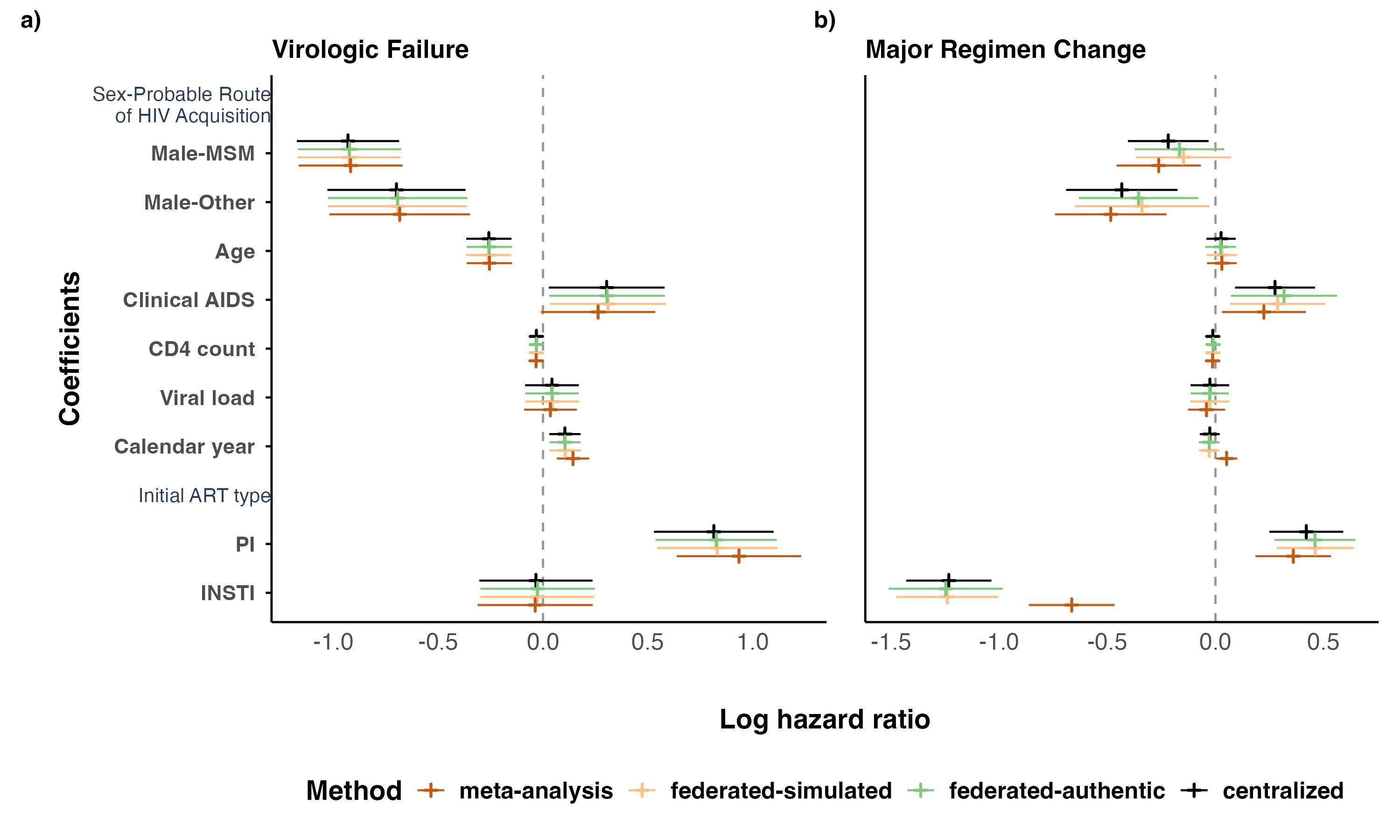}
\vspace{-2.5em}
\caption{Forest plots of results from adjusted analyses with multiple imputation for both virologic failure and major regimen change: for each coefficient and each analysis method, point estimate and corresponding 95\% CI are displayed; results for age reflect a 10-year increment; CD4 count is square-root transformed; viral load is $\log_{10}$-transformed; MSM, men who have sex with men; PI, (boosted) protease inhibitor; INSTI, integrase strand transfer inhibitor.}
\label{figure2}
\end{figure}

In terms of Wasserstein distance, the federated analysis by ODACH was superior to the meta-analysis (Table S1 and Table~\ref{table2}) in most situations, especially for the coefficients for initial regimens (PI and INSTI), calendar year, and clinical AIDS (as shown in Figure S1 and Figure~\ref{figure2}). For example, for the INSTI regimen coefficient in the multiply imputed major regimen change analysis, the meta-analysis produced a 95\% CI that did not overlap with that obtained from the centralized analysis, with which the federated analysis closely aligned.
Among the few exceptions where meta-analysis was closer to the centralized analysis than the federated analysis, the advantage was slim; e.g., the coefficient for CD4 count in the major regimen change analysis.
\par
In both the complete-case and multiply imputed analyses, there were very minor differences in coefficient estimates between the simulated and the authentic federated analyses, probably due to different computational environments. 
Despite these minor differences, the two versions of federated analyses aligned very well with each other (Figure S1 and Figure~\ref{figure2}).

\begin{table}[!ht]
\centering
\begin{threeparttable}
\caption{Baseline characteristics of individuals who had at least one available outcome in Brazil, Chile, and Mexico within CCASAnet in  and after 2015.}
\label{table1}
\begin{tabular}{lccc}
\toprule
 & \textbf{Brazil (n=1,928)} & \textbf{Chile (n=3,220)} & \textbf{Mexico (n=393)} \\
\midrule
\multicolumn{4}{l}{\textbf{Sex}}\\
\quad Female & 272 (14.1\%) & 375 (11.6\%) & 43 (10.9\%) \\
\quad Male   & 1{,}656 (85.9\%) & 2{,}784 (86.5\%) & 350 (89.1\%) \\
\quad Missing     & 0 (0\%) & 61 (1.9\%) & 0 (0\%) \\
\addlinespace
\multicolumn{4}{l}{\textbf{Age}}\\
\quad Median (IQR) & 31 (25, 40) & 31 (27, 38) & 32 (27, 42) \\
\addlinespace
\multicolumn{4}{l}{\textbf{Probable route of HIV acquisition}}\\
\quad Sexual & 519 (26.9\%) & 792 (24.6\%) & 80 (20.4\%) \\
\quad MSM    & 1{,}163 (60.3\%) & 2{,}399 (74.5\%) & 297 (75.6\%) \\
\quad Other  & 14 (0.7\%) & 7 (0.2\%) & 5 (1.3\%) \\
\quad Missing    & 232 (12.0\%) & 22 (0.7\%) & 11 (2.8\%) \\
\addlinespace
\multicolumn{4}{l}{\textbf{Clinical AIDS}}\\
\quad Not clinical AIDS & 1{,}485 (77.0\%) & 2{,}680 (83.2\%) & 313 (79.6\%) \\
\quad Clinical AIDS      & 443 (23.0\%) & 540 (16.8\%) & 80 (20.4\%) \\
\addlinespace
\multicolumn{4}{l}{\textbf{CD4 count}\textsuperscript{a}}\\
\quad Median (IQR) & 339 (110, 554) & 327 (169, 495) & 245 (81, 412) \\
\quad Missing & 384 (19.9\%) & 184 (5.7\%) & 47 (12.0\%) \\
\addlinespace
\multicolumn{4}{l}{\textbf{Viral load}\textsuperscript{b}}\\
\quad Median (IQR) & 4.62 (3.97, 5.27) & 4.84 (4.22, 5.40) & 5.11 (4.44, 5.65) \\
\quad Missing & 576 (29.9\%) & 500 (15.5\%) & 44 (11.2\%) \\
\addlinespace
\multicolumn{4}{l}{\textbf{Calendar year}}\\
\quad Median (IQR) & 2020 (2017, 2022) & 2018 (2017, 2020) & 2016 (2015, 2018) \\
\addlinespace
\multicolumn{4}{l}{\textbf{Initial ART}}\\
\quad NNRTI & 329 (17.1\%) & 967 (30.0\%) & 245 (62.3\%) \\
\quad PI    & 70 (3.6\%) & 416 (12.9\%) & 26 (6.6\%) \\
\quad INSTI & 1{,}529 (79.3\%) & 1{,}837 (57.0\%) & 122 (31.0\%) \\
\bottomrule
\bottomrule
\addlinespace
\multicolumn{4}{l}{\textbf{Virologic Failure (pooled sample size, n=4{,}670)}}\\
\quad Count of available outcomes & 1{,}802 & 2{,}488 & 380 \\
\quad Count of events & 190 & 192  & 38 \\
\addlinespace
\multicolumn{4}{l}{\textbf{Major Regimen Change (pooled sample size, n=5{,}541)}}\\
\quad Count of available outcomes & 1{,}928 & 3{,}220 & 393 \\
\quad Count of events & 211  & 498  & 215  \\
\bottomrule
\end{tabular}
\begin{tablenotes}
\footnotesize
\item \textit{Abbreviations:} MSM, men who have sex with men; NNRTI, non-nucleoside reverse transcriptase inhibitor; PI, (boosted) protease inhibitor; INSTI, integrase strand transfer inhibitor.
\item[a] The unit of CD4 count is cells/mm3.
\item[b] Viral load (copies/ml) is $\log_{10}$-transformed.
\end{tablenotes}
\end{threeparttable}
\end{table}

\begin{table}[!ht]
\centering
\begin{threeparttable}
\caption{Results of the adjusted analyses with multiple imputation: estimates and 95\% confidence intervals of hazard ratio are provided for the centralized analysis while Wasserstein distances are provided for the federated analyses by ODACH and the meta-analysis.}
\label{table2}
\small
\setlength{\tabcolsep}{6pt}

\begin{tabular}{
  l l                
  c c c              
}
\toprule
\multicolumn{5}{l}{\textbf{Virologic Failure}} \\
\cmidrule(lr){1-5}
\multirow{2}{*}{\textbf{Coefficient}} & \multirow{2}{*}{\textbf{Estimate (95\% CI)}\textsuperscript{b}} &
\multicolumn{3}{c}{\textbf{Wasserstein Distance}} \\
\cmidrule(lr){3-5}
& & \textbf{Meta} & \textbf{Federated.simulated}\textsuperscript{c} & \textbf{Federated.authentic}\textsuperscript{d} \\
\midrule

\multicolumn{5}{l}{\textbf{Sex--Probable route of HIV acquisition}} \\
\quad Female (reference) & 1.00 & & & \\
\quad Male--MSM    & 0.39 (0.31, 0.50) & 0.0126 & 0.0048 & 0.0075 \\
\quad Male--Other  & 0.50 (0.36, 0.69) & 0.0159 & 0.0058 & 0.0056 \\[0.4em]

\textbf{Age}         & 0.97 (0.96, 0.99) & 0.0003 & 0.0001 & 0.0002 \\[0.4em]

\textbf{Clinical AIDS} & 1.36 (1.03, 1.79) & 0.0411 & 0.0063 & 0.0020 \\[0.4em]

\textbf{CD4 count}\textsuperscript{a}  & 0.97 (0.95, 0.99) & 0.0015 & 0.0005 & 0.0002 \\[0.4em]

\textbf{Viral load}\textsuperscript{a} & 1.04 (0.92, 1.19) & 0.0076 & 0.0017 & 0.0009 \\[0.4em]

\textbf{Calendar year} & 1.11 (1.03, 1.20) & 0.0388 & 0.0017 & 0.0005 \\[0.4em]

\multicolumn{5}{l}{\textbf{Initial ART}} \\
\quad NNRTI (reference)         & 1.00  & & & \\
\quad PI            & 2.26 (1.70, 3.01) & 0.1200 & 0.0161 & 0.0114 \\
\quad INSTI         & 0.97 (0.74, 1.27) & 0.0044 & 0.0058 & 0.0083 \\

\midrule
\midrule
\addlinespace[0.6em]

\multicolumn{5}{l}{\textbf{Major Regimen Change}} \\
\cmidrule(lr){1-5}
\multirow{2}{*}{\textbf{Coefficient}} & \multirow{2}{*}{\textbf{Estimate (95\% CI)}\textsuperscript{b}} &
\multicolumn{3}{c}{\textbf{Wasserstein Distance}} \\
\cmidrule(lr){3-5}
& & \textbf{Meta} & \textbf{Federated.simulated}\textsuperscript{c} & \textbf{Federated.authentic}\textsuperscript{d} \\
\midrule

\multicolumn{5}{l}{\textbf{Sex--Probable route of HIV acquisition}} \\
\quad Female (reference) & 1.00 & & & \\
\quad Male--MSM    & 0.80 (0.67, 0.97) & 0.0444 & 0.0726 & 0.0531 \\
\quad Male--Other  & 0.65 (0.50, 0.84) & 0.0511 & 0.0972 & 0.0780 \\[0.4em]

\textbf{Age}         & 1.00 (1.00, 1.01) & 0.0004 & 0.0004 & 0.0003 \\[0.4em]

\textbf{Clinical AIDS} & 1.32 (1.09, 1.58) & 0.0516 & 0.0218 & 0.0520 \\[0.4em]

\textbf{CD4 count}\textsuperscript{a}  & 0.99 (0.98, 1.00) & 0.0008 & 0.0009 & 0.0019 \\[0.4em]

\textbf{Viral load}\textsuperscript{a} & 0.97 (0.89, 1.07) & 0.0157 & 0.0007 & 0.0013 \\[0.4em]

\textbf{Calendar year} & 0.97 (0.93, 1.02) & 0.0779 & 0.0024 & 0.0029 \\[0.4em]

\multicolumn{5}{l}{\textbf{Initial ART}} \\
\quad NNRTI (reference) & 1.00  & & & \\
\quad PI            & 1.52 (1.28, 1.81) & 0.0602 & 0.0410 & 0.0407 \\
\quad INSTI         & 0.29 (0.24, 0.35) & 0.5687 & 0.0207 & 0.0369 \\

\bottomrule
\end{tabular}

\vspace{0.3em}
\begin{tablenotes}
\footnotesize
\item \textit{Abbreviations:} CI, confidence interval; MSM, men who have sex with men; NNRTI, non-nucleoside reverse transcriptase inhibitor; PI, (boosted) protease inhibitor; INSTI, integrase strand transfer inhibitor.
\item[a] CD4 count (cells/mm3) is square-root transformed; viral load (copies/ml) is $\log_{10}$-transformed.
\item[b] Estimate (95\% CI) = hazard ratio estimate (95\% confidence interval) from centralized analysis.
\item[c] Federated.simulated = simulated federated analysis with Chile as the lead site.
\item[d] Federated.authentic = authentic federated analysis with Chile as the lead site.
\end{tablenotes}
\end{threeparttable}
\end{table}

\subsection*{Local Analyst Feedback}
Running the authentic federated analysis for each time-to-event endpoint took approximately 1.4 hours, covering both unadjusted and adjusted analysis. The total estimated average time in video calls was approximately 3.8 hours for local sites and 5.8 hours for VUMC analysts because of multiple separate calls with local sites. Analysts at the three sites reported spending an average of 13 total hours preparing for and participating in the federated analysis.

The local analysts uniformly found the analysis protocol, the file sharing mechanism, the practice sessions, and the communication with the VUMC coordinating center to be very helpful and effective. Two of the three analysts reported experiencing technical issues. These included issues related to downloading or applying needed packages and difficulties regarding the number of displayed decimal places for the intermediate outputs. These issues were ultimately resolved in coordination with the VUMC team during video calls. One analyst said that they would have liked to better understand the details of the analysis. One analyst felt that federated learning is a very trustworthy analysis approach moving forward, whereas the other two reported it as somewhat trustworthy. One analyst expressed concerns that data collection and data quality control were not standardized across sites. All analysts recognized the privacy-preserving and regulatory advantages offered by federated learning, and one analyst felt that it increased country-level autonomy while still enabling collaboration. However, all analysts pointed to limitations in increased logistics and coordination complexities. In conclusion, all analysts reported an overall positive experience and said they would be willing to participate in a future federated analysis.

\section*{Discussion}
This work demonstrates the feasibility of applying a federated statistical analysis framework in a multi-national observational HIV cohort. For two important time-to-event HIV endpoints, our empirical evidence in an authentic distributed environment supports statistical validity and operational readiness of the privacy-preserving and communication-efficient ODACH algorithm.  \cite{luo2022odach} Our authentic implementation of a federated analysis yielded estimates almost identical to the simulated federated analysis and closer to the centralized analysis than the meta-analysis. The positive experience reported by all participating analysts as well as their willingness to engage in similar analyses in the future suggest that federated analysis is a viable approach for circumventing barriers to IPD sharing. 
To our knowledge, this represents the first authentic implementation of federated learning in a multinational observational HIV cohort study setting.


A key challenge in real-world federated statistical analysis is handling missing data, which remains underexplored in practical federated implementations. We addressed this by integrating multiple imputation into the federated workflow in a sequential manner. Specifically, each site performed imputation locally without sharing information across sites, after which the federated analysis was run on each imputed dataset and results were pooled. 
In this regard, the number of imputed datasets determined the number of communication rounds (uploads and downloads), because each imputed dataset underwent its own federated analysis.
Since this information exchange process can be time-consuming and prone to errors, we limited our study to ten imputations as a compromise between controlling random variability across imputations and reducing the operational burden.


Apart from federated Cox proportional hazards regression (e.g., ODACH), there are other communication-efficient federated algorithms for survival analysis and many other statistical models. A heterogeneity-aware federated algorithm for the Fine-Gray model, \cite{Fine1999} which is often used to estimate associations between risk factors and cumulative incidences of a time-to-event outcome in the presence of competing events,
has been proposed\cite{zhang2024oneshot} but is not ready for implementation.
For many common statistical models, there are available federated algorithms within the PDA (privacy-preserving distributed algorithms) framework,\cite{pda2026} including for logistic regression, \cite{duan2020odal} Cox proportional hazards models, \cite{duan2020odac,luo2022odach,liang2025federated} and linear mixed effects models. \cite{luo2022dlmm} Continued efforts to develop federated algorithms for novel statistical models are warranted.

Despite the advantages of federated analysis demonstrated here, there are certainly some analysis tasks that are difficult to perform with federated analyses. For example, model diagnostics are often performed in centralized analyses, but it is challenging to conduct these in a federated framework. Specifically, our analyses made linearity assumptions for age, square-root-transformed CD4 count, $\log_{10}$-transformed viral load, and calendar year, but checking these assumptions in a distributed environment is not an easy task. In addition, sensitivity analyses or simply updating analyses in response to a review or new data requires assembling the local analysts again, thus adding to the complexity and burden of completing authentic federated analyses in a timely manner.

Beyond methodological accuracy lies a distinct set of challenges that determines real-world success. Statistically valid and privacy-preserving federated algorithms do not immediately translate to their successful implementation in the real world, where data resources, infrastructure, human networks, and cross-site coordination play critical roles. \cite{peltonen2026networks,loftus2022fedMIrealworld} We learned these from our own experience. First, authentic implementation by three separate analysts in three different countries required multiple video conferences to ensure that the computational environments were correctly configured  and software packages properly installed. Scaling this to a larger number of sites or countries (e.g., to also include sites in Africa) would have required substantially greater coordinating effort, with language differences across countries potentially introducing an additional layer of complexity. Second, we were also in an idealized setting, where the data coordinating center at VUMC had already harmonized data from the sites on which we wrote code. Our investigation would have been much more challenging without centralized and harmonized data. Finally, infrastructure for federated analysis with multiple imputation is lacking. The PDA-OTA,\cite{pdaota2026} a web-based platform, is designed for information exchange for authentic federated analysis, accommodating ODACH. Instead of using institution-authorized cloud services, the platform supports asynchronous file uploads and downloads, and requires analysts to run R code on local computers.
One key limitation of this platform is that it cannot easily handle local multiple imputation in the sense that it would require multiple independent project proposals for different sets of locally imputed datasets, and thus we did not use it.



With these opportunities and limitations noted, given the increased difficulties in sharing multi-national data, particularly from PWH, federated analysis—a collaborative framework for analyzing distributed data without centralizing it—is poised for broader adoption in the years ahead.



{\small
\bibliographystyle{ama}
\bibliography{refs}
}

\end{CJK}\end{document}